\documentstyle[prl,aps]{revtex}
\large
\begin{document}
\twocolumn

\draft

\title{ A Modified Approach to Single-Spin Detection\\
Using Magnetic Resonance Force Microscopy}

\author{{ Gennady P. Berman$^1$, and Vladimir I. Tsifrinovich$^2$}}

\address{
$^1$Theoretical Division and CNLS, Los Alamos National Laboratory, 
Los Alamos, New Mexico 87545\\
$^2$Department of Applied Mathematics and Physics, Polytechnic University,
Six Metrotech Center, Brooklyn NY 11201}

\maketitle

\begin{abstract} 
The magnetic moment of a single spin interacting with a cantilever in magnetic resonance force microscopy (MRFM) experiences quantum jumps in orientation rather than smooth oscillations. These jumps cannot be detected by a conventional MRFM based on observation of driven resonant oscillations of a cantilever. In this paper, we propose a method which will allow detection
of the magnetic signal from a single spin using a modification of a conventional MRFM. We estimate the opportunity to detect the magnetic signal from a single proton.     
\newline
\renewcommand{\baselinestretch}{1.656}

%
\end{abstract}
{\bf 1. Introduction}\\ \ \\
In 1991, Sidles \cite{1} proposed a force detection of nuclear magnetic resonance which could achieve a sensitivity to detect the magnetic signal from a single proton spin. This idea gave birth to magnetic resonance force microscopy (MRFM). MRFM  has been used to increase the sensitivity and spatial resolution for an electron spin resonance \cite{2}, ferromagnetic resonance \cite{3}, and nuclear magnetic resonance \cite{4}. In conventional MRFM, a magnetic particle produces a non-uniform magnetic field which attracts or repels the magnetic moment of a sample placed on a cantilever, depending on the direction of a magnetic moment. The magnetic resonance technique provides oscillations of the magnetic moment orientation which cause resonant vibrations of the cantilever. These vibrations can be detected, for example, using optical methods. This method clearly does not apply to single-spin detection, because when interacting with a cantilever a single spin exhibits quantum jumps rather than smooth oscillations of its $z$-component (a component in the direction of the permanent magnetic field). 
Indeed, observation of driven oscillations of a cantilever would have allowed one to measure a continuous change of a single spin $z$-component, a measurement which is forbidden by quantum mechanics. 

In this paper, we suggest a modification of the conventional MRFM technique which will allow one to detect a single spin. The main idea is the following. While a single spin located on the ``classical'' cantilever and driven by the external resonant radio-frequency field is not able to produce a harmonic magnetic force (because of quantum jumps), it can still produce a periodic magnetic force (with a period of the cantilever oscillations) caused by a periodic sequence of radio-frequency pulses. In our scheme, a periodic sequence of short pulses produces periodic spin flips. The time to flip the spin is much shorter than the period of the cantilever oscillations. Thus, the cantilever does not destroy each individual spin flip. The periodic magnetic force produced by these spin flips has many harmonics including one which is resonant with the cantilever oscillations. Thus, this pulsed periodic force can drive the resonant vibrations of the cantilever as well as a harmonic force. The whole picture reminds two children, which drive the resonant vibrations of a swing pushing it periodically back and forth during short time-intervals (see Fig. 1). (The difference is that in the case of a cantilever the magnetic force changes the direction during short time-intervals, but does not vanish between the successive pulses.) In Section 2, we describe the dynamics of a cantilever. In Section 3, we explain our strategy for detection of a single spin. In Section 4, we estimate the value of the magnetic signal from a single proton.\\ \ \\
{\bf 2. Dynamics of the cantilever}\\ \ \\
Fig. 2, shows the typical geometry of a MRFM experiment. The sample, $S$, containing the spin to be measured is mounted at the free end of the cantilever. A small ferromagnetic particle, $F$, produces  non-uniform magnetic field, $B_f$, in the sample, $S$. A large uniform magnetic field, $B_0$, is directed in the positive $z$-direction, while the radio-frequency ({\it rf}) field, $B_1$, is polarized in the $xy$-plane. The non-uniform magnetic field produced by the ferromagnetic probe changes the magnetic field in the sample in such a way that only selected spins are in resonance with the {\it rf} field. The change in orientation of selected spins
under the action of the {\it rf} field influences the magnetic force between the ferromagnetic particle and the sample, which causes the oscillations of the cantilever.
For optimal MRFM imaging (to scan a sample of an arbitrary form) it is desirable to place the ferromagnetic particle on a cantilever. However, this replacement faces technical problems. 

Assume the cantilever has dimensions: $l_c$ (length), $w_c$
(width), and $t_c$ (thickness) (see Fig. 3). The equation of motion for the cantilever displacement, $z_c(x,t)$, under the action of the external harmonic force, $F_\omega\exp(i\omega t)$, can be written as \cite{5},
$$
{{Et^2_c}\over{12}}{{\partial^4z_c}\over{\partial x^4}}+\rho {{\partial^2z_c}\over{\partial t^2}}={{1}\over{t_cw_c}}\delta(x_c-l_c)F_\omega\exp(i\omega t).\eqno(1)
$$
Here, $E$ is Young's modulus, $\rho$ is the density of a cantilever, and the external force is polarized along the $z$-axes and is applied to the free end of a cantilever. With boundary conditions,
$$
z_c|_{x=0}={{\partial z_c}\over\partial x}|_{x=0}={{\partial^2 z_c}\over\partial x^2}|_{x=l_c}={{\partial^3 z_c}\over\partial x^3}|_{x=l_c}=0,\eqno(2)
$$
the solution of Eq. (1) for driven oscillations for the point $z_c(x=l_c,t)$ takes the form,
$$
z_c={{4}\over{m}}\sum_{n=1}^\infty{{F_\omega}\over{\omega^2_n-\omega^2}} e^{i\omega t},\eqno(3)
$$
where $m$ is the mass of a cantilever. The summation in Eq. (3) is taken over all eigenfrequencies of the cantilever,
$$
\omega_n=(t_c\lambda^2_n/l^2_c)(E/12\rho)^{1/2},\eqno(4)
$$
where the coefficients, $\lambda_n$, satisfy the equation,
$$
\cosh\lambda_n\cos\lambda_n=-1.\eqno(5)
$$
Neglecting all terms in Eq. (3) except for the first one with $n=1$, and taking into consideration the finite value of the quality factor, $Q$, of the cantilever, we have from Eq. (3),
$$
z_c\approx{{4F_\omega/m}\over{\omega^2_c-\omega^2+i\omega^2/Q}}e^{i\omega t},\eqno(6)
$$
where $\omega_c$ is the lowest eigenfrequency of the cantilever,
$$
\omega_c=\omega_1\approx 1.04(t_c/l^2_c)(E/\rho)^{1/2},\eqno(7)
$$
and $\lambda_1\approx 1.9$.

For $\omega=0$, this expression transforms into,
$$
F_0\approx {{m\omega_c^2}\over{4}}z_c,~ or~ z_c\approx F_0/k_c.\eqno(8)
$$
The coefficient, $k_c=m\omega^2_c/4$, is the effective ``spring constant'' of the cantilever. At resonance, $\omega=\omega_c$, Eq. (6) takes the form,
$$
z_c\approx -iQ(F_{\omega}/k_c)e^{i\omega_c t}.\eqno(9)
$$
Eq. (9) describes the resonant enhancement of the oscillations by a quality factor, $Q$. 

Using the fluctuation-dissipation theorem \cite{6}, one can estimate both the root-mean-square (rms) vibration amplitude, $z_{rms}$, and the $rms$ force, $F_{rms}$, caused by the thermal vibrations,
$$
z_{rms}=\Bigg[2\hbar Im(\chi) \coth\Bigg({{\hbar\omega}\over{2k_BT}}\Bigg)\Delta f\Bigg]^{1/2}\approx \eqno(10)
$$
$$
[4k_BTQ\Delta f/k_c\omega_c]^{1/2},
$$
$$
F_{rms}=\Bigg[{{2\hbar Im({\chi})}\over{|\chi|^2}}\coth\Bigg({{\hbar\omega}\over{2k_BT}}\Bigg)\Delta f\Bigg]^{1/2}\approx
$$
$$
\Bigg[{{ 4k_BTk_c\Delta f}\over{Q\omega_c}}\Bigg]^{1/2}=(k_c/Q)z_{rms},
$$
where $\chi=iQ/k_c$ is the resonant susceptibility, and $\Delta f$ is the noise bandwidth of the cantilever. The second expression in Eq. (10) determines the minimum detectable $rms$ force.

At present, the driven cantilever vibrations in MRFM are caused by the harmonic oscillations of the $z$-component of a magnetic moment of a sample. As an example, in the first experiment \cite{4} on the proton magnetic resonance in ammonium nitrate, the frequency of the $rf$ field, $B_1$, was modulated with a cantilever frequency, $\omega_c$. Then, the effective magnetic field which acts
on a proton in the rotating reference frame changed adiabatically its direction (cyclic adiabatic inversion). As a result, the $z$-component of a nuclear magnetic moment in the sample oscillates with the frequency $\omega_c$, producing a resonant force on the cantilever. In this experiment, the permanent magnetic field was, $B_0\approx 2.35$T, the amplitude of the $rf$ field was, $B_1\sim 10^{-3}$T. These parameters correspond to the NMR frequency,
$$
f_0=(\gamma/2\pi)B_0=100MHz,
$$
and the Rabi frequency,
$$
f_R=(\gamma/2\pi)B_1\sim 42.6kHz,
$$
where $\gamma$ is the proton gyromagnetic ratio ($\gamma/2\pi=4.26\times 10^7$Hz/T). The modulated frequency of the $rf$ field,
$$
f=f_0+(\Omega/2\pi)\sin\omega_c t,
$$
had the peak deviation: $\Omega/2\pi\sim 100$kHz. An iron particle created the $rms$ magnetic force of $1.1\times 10^{-14}$N. A 900$\stackrel{o}{A}$ thick silicon nitride cantilever had the spring constant, $k_c=10^{-3}$N/m, resonant frequency, $\omega_c/2\pi=1.4$kHz, and a noise bandwidth, $\Delta f=0.25$Hz.
The $rms$ vibration noise at room temperature was: $z_{rms}\approx 5\stackrel{o}{A}$, and the corresponding value of the signal was $z_{c,rms}\approx 110\stackrel{o}{A}$, which provided a signal-to-noise ratio of 22. Consequently, in this experiment, the minimum detectable {\it rms} force was $5\times 10^{-16}$N.\\ \ \\
{\bf 2. Detection of a single spin}\\ \ \\
Currently, MRFM exploits resonant vibrations of a cantilever driven by the continuous oscillations of the $z$-component of the sample magnetic moment, ${m}_z$. This method does not fit for a single spin detection. Indeed, a cantilever interacts with a spin like a macroscopic measuring device interacts with a quantum system. A single spin 1/2 interacting with a cantilever can exhibit only quantum jumps between the ground state and the excited state, but not a continuous change of $m_z$ between the values $-\hbar/2$ and $\hbar/2$. As a result, a cantilever will experience a magnetic force, $F_z=\pm F=\pm m_0\nabla B_z$, which randomly changes its sign ($m_0=\gamma\hbar/2$ is the magnetic moment of a particle). Random external force crucially reduces the resonant susceptibility of a cantilever.

To provide a resonant response in MRFM for single spin detection we suggest the following modification of the conventional MRFM method. One should change the direction of the spin during the time interval which is short in comparison with the period of the cantilever oscillations, $T_c=2\pi/\omega_c$. For a cantilever such change in spin direction is equivalent to a quantum jump. 

To change the direction of a spin,  one can use either a powerful $\pi$-pulse or fast adiabatic inversion. Consider, for example, the first opportunity. Then, one applies a $rf$ pulse with a NMR frequency, $\gamma (B_0+B_f)$, the amplitude, $\pi/\tau$, and a duration, $\tau$ which is small in comparison with the period of a cantilever vibration, $T_c$. During a short time interval, $\tau$, the wave function of the spin evolves according to the Schr\"odinger equation, {\it i.e.} during the time interval, $\tau$, spin transfers from the ground state to the excited state or vice versa. Next, one applies the same $\pi$-pulses periodically, with a period  $T_c/2$. In this case, the cantilever experiences a periodic force, $F(t)=F(t+T_c)=\pm m_0\nabla B_z$ which induces resonant vibrations of a cantilever. Let's choose the even function $F(t)$ (see Fig. 4),
$$
F(t)=\cases 
{F,& $-T_c/4<t<T_c/4$,\cr
-F,& $-T_c/2<t<-T_c/4$, ~or~ $T_c/4<t<T_c/2$.\cr
}\eqno(11)
$$
The Fourier component of $F(t)$ on the cantilever's frequency, $\omega_c$, is: $F_\omega\cos\omega_c t$, where the amplitude, $F_\omega$, is (below, to simplify notation, we use $F_\omega$ instead of $F_{\omega_c}$),
$$
F_\omega={{2}\over{T_c}}\int^{T_c/2}_{-T_c/2}F(t)\cos\omega_c tdt=4F/\pi.\eqno(12)
$$
This component drives the resonant vibrations of a cantilever.\\ \ \\
{\bf 3. A signal from a single proton spin}\\ \ \\
One of the most important problems of MRFM is the detection of a single proton spin. 
Now we estimate the opportunity to detect a single proton spin using a method described in section 2. The gradient of the magnetic field produced by a spherical ferromagnetic tip is given by the expression: $|\partial B_z/\partial z|=2\mu_0Mr^3/(r+d)^4$, where $\mu_0$ is a permeability of the free space ($\mu_0=4\pi\times 10^{-7}$H/m), $M$ is a magnetization of a tip, 
 $r$ is its  radius, and $d$ is a separation 
 between the tip and a spin \cite{7}. Assume that the iron tip of the 
radius $r= 300\stackrel{0}{A}$ is placed at $d=100\stackrel{0}{A}$ from a spin. It provides the gradient $|\nabla B_z|\approx 4.5\times 10^7$T/m. 
The magnetic moment of a proton is: $m_0\approx 1.4\times 10^{-26}$J/T. For the gradient, $4.5\times 10^7$T/m, we get the magnetic force: $F=6.3\times 10^{-19}$N. The corresponding value of $F_\omega$ in Eq. (12) is: $8.0\times 10^{-19}$N.

To estimate the effect produced by this force, we consider, as an example, an ultrathin silicon cantilever reported in \cite{8}. It had a thickness, $t_c=600\stackrel{o}{A}$, and a length, $l_c=0.22\mu m$. Substituting these values in Eq. (7)  and taking the Young's modulus, $E \approx 1.6\times 10^{11}N/m^2$, and the density, $\rho=2.33\times 10^3kg/m^3$, one derives the experimental value of the cantilever's frequency: $\omega_c/2\pi\approx 1.7$kHz. Other reported in \cite{8} quantities are: the spring constant, $k_c=6.5\times 10^{-6}$N/m, the quality factor, $Q=6700$, the noise bandwidth $\Delta f=0.4$Hz.

In experiments \cite{8}, the electrostatic test force,
$F_{\omega,rms}\approx 3.6\times 10^{-17}$N, produced the vibration amplitude corresponding to Eq. (9),
$$
z_{c,rms}=(Q/k_c)F_{\omega,rms}\approx 370\stackrel{o}{A}.\eqno(13)
$$
The noise level observed at room temperature ($T=T_r=300$K) corresponds to the value estimated from Eq. (10): $z_{rms}(T_r)\approx 250\stackrel{o}{A}$. Thus, the minimum detectable force at room temperature was,
$$
F_{rms}(T_r)=(k_c/Q)z_{rms}(T_r)\approx 2.4\times 10^{-17}N.\eqno(14)
$$

The $rms$ force produced by a single proton is,
$$
F_{\omega,rms}=F_\omega/\sqrt{2}\approx 5.7\times 10^{-19}N.\eqno(15)
$$
A cantilever described in \cite{8} could detect such a force at temperature,
$$
T<300(F_{\omega,rms}/F_{rms}(T_r))^2\approx 0.17K,\eqno(16)
$$
as $F_{rms}\sim T^{1/2}$.

Note, that the induced periodic inversion of a single proton spin assumes that the probability of spontaneous inversion, $W_s$, is negligible during the time of the experiment. On the other hand, driven oscillations of a cantilever can be observed at times: $t>2Q/\omega_c$. This imposes a restriction on the
probability, $W_s$: $W^{-1}_s>2Q/\omega_c$. For parameters in \cite{8}, it gives: $W^{-1}_s>1$s.

Now we shall discuss a possible improvement of cantilever's parameters which make the opportunity of a single proton spin detection more feasible. First, we rewrite the expression for the spring constant taking into consideration Eq. (7),
$$
k_c=(\lambda^4_1/48)w_c E(t_c/l_c)^3.\eqno(17)
$$
Combining the minimum values of the thickness and the width with the maximum value of the length reported in \cite{8},
$$
t_{min}=500\stackrel{o}{A},~l_{max}=0.40mm,~w_{min}=4\mu m,\eqno(18)
$$
we get: $k_c=3.4\times 10^{-7}$N/m, and $f_c=0.43$kHz. Now, it seems to be reasonable to expect the increase of a cantilever's quality factor to $10^5$ and the gradient of magnetic field to $10^8$T/m. For such combination of parameters, we have the magnetic force: $F_{\omega,rms}\approx 1.3\times 10^{-18}$N, and the corresponding vibration amplitude: $z_{c,rms}\approx 3800\stackrel{o}{A}$. Assuming that the noise bandwidth, $\Delta f$, is proportional to $f_c/Q$, we get: $\Delta f\approx 6.8\times 10^{-3}$Hz. Thus, the noise level at room temperature will be: $z_{rms}(T_r)\approx 1100\stackrel{o}{A}$, and the minimum detectable force: $F_{rms}(T_r)\approx 3.7\times 10^{-19}$N. In this case, a single proton spin can be detected even at room temperature.\\ \ \\
{\bf Conclusion}\\ \ \\
In this paper, we discuss the opportunities of a single spin detection using
MRFM. We point out that conventional MRFM methods cannot perform a 
single-spin measurement. They are based on detection of the resonant vibrations
of a cantilever driven by the oscillating magnetic moment of a sample. For
a single spin interacting with a measurement device, one faces quantum
jumps rather than oscillating magnetic moment. 
We propose modification of the traditional MRFM which allows a 
single-spin measurement. The main idea is that under the action of a short {\it rf}
pulse, the spin changes its orientation in a time interval small compared with the
period of the cantilever. During this small time interval one can consider the spin as a pure quantum system in an
external field. Next, one applies a periodic sequence of {\it rf} pulses. If the
time interval between {\it rf} pulses is equal to a half the period of the
cantilever, one can observe the resonant vibrations of a cantilever as
well as they were produced by the oscillating magnetic moment.
We have estimated the possibility of detecting a single proton spin using the
method suggested in this paper. For an ultrathin cantilever with parameters reported in \cite{8}, a
proton spin could be detected at temperatures below 0.17K. Expected
improvements of cantilever parameters promise to increase the temperature region up to room temperatures.\\ \ \\
{\bf Acknowledgments}\\ \ \\
We thank P.C. Hammel and G.D. Doolen for discussions. This work  was supported by the Department of Energy under contract W-7405-ENG-36, and by the National Security Agency.\\ \ \\
{\bf Figure captions}\\ \ \\
Fig. 1:~Driving the resonant vibrations of a swing pushing it periodically back and forth during short time-intervals.\\ \ \\
Fig. 2:~Schematic experimental set-up of MRFM: $B_0$-the permanent magnetic field; $B_1$-the radio-frequency ({\it rf}) field; $B_f$-non-uniform magnetic field produced by a ferromagnetic particle, $F$, in the sample, $S$.\\ \ \\
Fig. 3:~Dimensions of a cantilever: $l_c$, $w_c$, and $t_c$ -- the length, width, and thickness of a cantilever.\\ \ \\
Fig. 4:~ Periodic force, $F(t)$, generated by a single spin.
\end{document}